\documentclass[twocolumn,amsmath,amssymb]{revtex4}

\usepackage{psfrag}
\usepackage[dvips]{graphicx}
\usepackage{feynmp}
\usepackage{multirow}

\begin{document}

\title{A Twisted Custodial Symmetry in the Two-Higgs-Doublet Model}

\author{J.-M. G\'erard}
\email[]{gerard@fyma.ucl.ac.be}
\affiliation{Institut de Physique Th\'eorique and Centre for Particle Physics and Phenomenology (CP3)\\
Universit\'e Catholique de Louvain, Belgium}

\author{M. Herquet}
\email[]{mherquet@fyma.ucl.ac.be}
\affiliation{Institut de Physique Th\'eorique and Centre for Particle Physics and Phenomenology (CP3)\\
Universit\'e Catholique de Louvain, Belgium}

\preprint{CP3-07-07}

\date{March 5, 2007}

\begin{abstract}
In the Standard Model for electroweak interactions, the Higgs sector is known to display a ``custodial'' symmetry protecting the mass relation $m_{W^\pm}^2=m_{W_3}^2$ from large corrections. When considering extensions of the scalar sector, this symmetry has to be introduced by hand in order to pass current electroweak precision tests in a natural way. In this Letter we implement a generalized custodial symmetry in the two-Higgs-doublet model. Assuming the invariance of the potential under $CP$ transformations, we prove the existence of a new custodial scenario characterized by $m_{H^\pm}^2=m_{H^0}^2$ instead of $m_{H^\pm}^2=m_{A^0}^2$. Consequently the pseudoscalar $A^0$ may be much lighter than the charged $H^\pm$, giving rise to interesting phenomenology.
\end{abstract}
\maketitle

\section{Introduction}
 In the Standard Model (SM) for electroweak interactions, the spontaneous symmetry breaking mechanism is known to have important phenomenological consequences on the bosonic sector of the theory. They can be inferred \cite{Veltman:1997nm} from the high degree of symmetry of the most general renormalizable scalar potential built for one Higgs doublet transforming under the local $SU(2)_L\times U(1)_Y$. Gauge invariance implies indeed an accidental $SO(4)$ symmetry acting upon the four components of the complex doublet. Through the Higgs mechanism, this global symmetry is spontaneously broken into $SO(3)$ under which the triplet $(\pi^\pm,\pi_3)$ of Goldstone bosons transforms. However this $SO(3)$ symmetry is explicitely broken by the electroweak gauge couplings $g_L$ and $g_Y$. In particular, the mass relation
\begin{equation}
\label{massrelsm}
 m_{W^\pm}^2=m_{Z^0}^2\left(\frac{g_L^2}{g_L^2+g_Y^2}\right)
\end{equation}
tells us that a massive triplet of vector bosons is only recovered in the limit of vanishing $g_Y$. A massless triplet including the charged $W^\pm$ and the photon can also form, but in the even less realistic limit of vanishing $g_L$. Yet, the $SO(3)$ symmetry of the Higgs potential is called ``custodial'' \cite{Sikivie:1980hm} since it protects the relation $m_{W^\pm}^2=m_{W_3}^2$ against loop corrections quadratic in the Higgs boson mass. These corrections might indeed conflict with the well measured value of the $\rho$-parameter.

In the Minimal Supersymmetric extension of the Standard Model (MSSM), one additional Higgs doublet is required in order to cancel the gauge anomalies induced by the fermionic superpartners. This implies the existence of five spin-zero physical states: a charged pair $H^\pm$ and three neutral ones. Given that the MSSM scalar potential is $CP$-invariant, the $h^0$ and  $H^0$ are defined to be the scalars while $A^0$ is the pseudoscalar of the theory. From a phenomenological point of view, this $CP$ assignment allows the $ZZh^0$ and $ZZH^0$ vertices but forbids the $ZZA^0$ one at the classical level. The general MSSM scalar potential is however not invariant under the custodial symmetry due to the presence of a $D$-term proportional to $g_L^2$. This gauge term lifts degeneracy of the $H^\pm$ and $A^0$ states, as can be seen from the tree-level mass relation
\begin{equation}
 m_{H^\pm}^2=m_{A^0}^2+m_{W^\pm}^2.
\end{equation}
Consequently the custodial $SO(3)$ symmetry with its distinctive degenerate mass spectrum is restored either in the standard decoupling limit for $A^0$ and $H^\pm$ (see for example \cite{Drees:1990dx}) or in the unphysical limit where the left-handed gauge interactions are switched off (i.e., $g_L\rightarrow 0$).

In the two-Higgs-doublet model (2HDM), such limits are circumvented since the scalar potential does not depend on the electroweak gauge couplings. An explicit calculation \cite{Toussaint:1978zm} of the one-loop corrections to $m_{W^\pm}^2=m_{W_3}^2$ in this model has shown that contributions quadratic in the Higgs bosons masses compensate each other in the limit where 
\begin{equation}
 m_{H^\pm}^2=m_{A^0}^2,
\end{equation}
namely if the charged $H^\pm$ and the neutral \textit{pseudoscalar} behave as a triplet under the custodial $SO(3)$. Surprisingly, it has been noted \cite{Chankowski:2000an} in the context of a rather peculiar $CP$-conserving 2HDM that these contributions could also cancel when 
\begin{equation}
 m_{H^\pm}^2=m_{H^0}^2 .
\end{equation}
In this case the mass degeneracy occurs between the charged $H^\pm$ and one neutral \textit{scalar}. The purpose of this Letter is to show that this second scenario can be implemented in a natural way within a generalized custodial symmetry. 

\section{Generalized Custodial Symmetry}
Consider the 2HDM based on  two $SU(2)_L$ doublets $\phi_1$ and $\phi_2$ with hypercharge $Y=+1$. Gauge invariance allows us to define four independent Hermitian operators
\begin{eqnarray}
\hat{A}&=& \phi_1^\dag \phi_1\nonumber\\
\hat{B}&=& \phi_2^\dag \phi_2\nonumber\\
\hat{C}&=& \Re\left(\phi_1^\dag\phi_2\right)=\frac{1}{2}\left(\phi_1^\dag\phi_2+\phi_2^\dag\phi_1\right)\nonumber\\
\hat{D}&=& \Im\left(\phi_1^\dag\phi_2\right)=-\frac{i}{2}\left(\phi_1^\dag\phi_2-\phi_2^\dag\phi_1\right)
\label{operators}
\end{eqnarray}
such that the most general scalar potential contains four linear and ten quadratic terms in $\hat{A}$, $\hat{B}$, $\hat{C}$ and $\hat{D}$. Using the well-known reparametrization freedom for $(\phi_1,\phi_2)$ \cite{Branco:1999fs,Davidson:2005cw}, we can assume without loss of generality to be in the so-called ``Higgs basis'' where only $\phi_1$ gets a nonzero vacuum expectation value (vev):
\begin{equation}
\label{HB}
 \langle\phi_1^0\rangle=v\quad\mathrm{and}\quad\langle\phi_2^0\rangle=0.
\end{equation}
In the SM, charge conservation is a direct consequence of the accidental $SO(4)$ symmetry. Here, charge conservation has to be assumed and an $SO(4)$ symmetry imposed. This global symmetry acting on the real components of
\begin{equation}
 \phi_1\equiv\frac{1}{\sqrt{2}}
\left(
\begin{array}{c}
\pi_1+i\pi_2 \\
\sigma_0+i\pi_3
\end{array}\right)
\end{equation}
is isomorphic to $SU(2)_L\times SU(2)_R/\mathbb{Z}_2$. The $SU(2)_L\times SU(2)_R$ chiral symmetry acts on the $[1/2,1/2]$ representation $M_1$ of the Higgs doublet $\phi_1$
\begin{equation}
M_1\equiv\frac{1}{\sqrt{2}}(\sigma_0\mathbb{I}+i\pi_a\tau^a)
\end{equation}
as
\begin{equation}
\label{transf1}
 M_1\rightarrow U_L M_1 U_R^\dag
\end{equation}
while $\mathbb{Z}_2$ is the discrete symmetry associated with the simultaneous change of sign of both left and right unitary matrices $U_{L,R}$. As explicitly demonstrated in \cite{Lytel:1980zh}, the invariance of the vacuum under the diagonal subgroup $SU(2)_{L+R}$ is necessary to ensure that relation $m_{W^\pm}^2=m_{W_3}^2$ does not suffer from large (i.e., quadratic in the Higgs bosons masses) corrections at the one-loop level. This vectorlike subgroup is obviously isomorphic to the custodial $SO(3)$ group. However, at this stage the chiral transformation for the $[1/2,1/2]$ representation $M_2$ of $\phi_2$ is not yet completely fixed. Indeed, only $SU(2)_L\times U(1)_Y$ is a local symmetry of the Lagrangian. For the bosonic sector of the theory, the conserved electric charge turns out to be $Q=T_L^3+T_R^3$ with $T^3_R$ the diagonal generator of the global $SU(2)_R$. So we still have the freedom to impose the invariance under
\begin{equation}
\label{transf2}
 M_2\rightarrow U_L M_2 V_R^\dag
\end{equation}
with
\begin{equation}
 V_R=X^\dag U_R X
\end{equation}
if the two-by-two unitary matrix $X$ commutes with $\exp(iT^3_R)$, namely
\begin{equation}
 X=\left(
\begin{array}{cc}
\exp(i\frac{\gamma}{2}) & 0 \\
0 & \exp(-i\frac{\gamma}{2})
\end{array}\right).
\end{equation}

It is straightforward to see that both $\hat{A}$ and $\hat{B}$ operators are invariant under the chiral transformations (\ref{transf1}) and (\ref{transf2}) while $\hat{C}$ and $\hat{D}$ are not if $\gamma$ is an arbitrary parameter. Nevertheless the linear combination
\begin{eqnarray}
\label{Cprime}
 \hat{C}'&\equiv& \frac{1}{2}\mathrm{Tr}(M_1 X M_2^\dag)=\frac{1}{2}\mathrm{Tr}(M_2 X^\dag M_1^\dag)\nonumber \\
&=&\cos(\frac{\gamma}{2})\hat{C}+\sin(\frac{\gamma}{2})\hat{D}
\end{eqnarray}
is always invariant, no matter the value of $\gamma$. Therefore, the most general custodial-invariant potential only contains three linear and six quadratic terms in $\hat{A}$, $\hat{B}$ and $\hat{C}'$:
\begin{eqnarray}
\label{potcust}
 V&=& -m_1 \hat{A}-m_2 \hat{B}-m_3\hat{C}'+\Lambda_1\hat{A}^2+\Lambda_2\hat{B}^2+\Lambda_3\hat{C}'^2\nonumber \\
& &+\Lambda_4\hat{A}\hat{B}+\Lambda_5\hat{A}\hat{C}'+\Lambda_6\hat{B}\hat{C}'.
\end{eqnarray}
The minimization conditions are easily derived to be
\begin{equation}
 m_1=\Lambda_1 v^2\quad \mathrm{and}\quad m_3=\frac{\Lambda_5}{2} v^2.
\end{equation}
We shall use these relations to substitute $\Lambda_1$ and $\Lambda_5$ for $m_1$ and $m_3$, respectively. 

The squared mass of $\phi_2^\pm$ is given by
\begin{equation}
 m_{H^\pm}^2=\frac{\Lambda_4}{2} v^2-m_2.
\end{equation}
A suitable $\gamma/2$ rotation allows us to reduce the full three-by-three mass matrix for the neutral fields into a single mass term 
\begin{equation}
\label{degmass}
 m_{H_3}^2=m_{H^\pm}^2
\end{equation}
for the state $H_3\equiv-\sin(\frac{\gamma}{2})\Re(\phi_2^0)+\cos(\frac{\gamma}{2})\Im(\phi_2^0)$ and a two-by-two mass matrix 
\begin{equation}
\label{matmass}
 \mathcal{M}^2=\left(\begin{array}{cc}
 2\Lambda_1 v^2 & \frac{\Lambda_5}{2} v^2 \\
 \frac{\Lambda_5}{2} v^2 & m_{H^\pm}^2+\frac{\Lambda_3}{2} v^2
\end{array}\right)
\end{equation}
for $(H_1,H_2)\equiv(\Re(\phi_1^0),\cos(\frac{\gamma}{2})\Re(\phi_2^0)+\sin(\frac{\gamma}{2})\Im(\phi_2^0))$. The $H_3$ is thus degenerate with $H^\pm$ in a triplet of $SO(3)$,  a clear signature of the custodial character of the potential (\ref{potcust}). The $H_{1,2}$ are singlets under this symmetry but mix if $\Lambda_5\neq 0$. In order to identify these neutral states in terms of the usual $CP$ eigenstates $h^0$, $H^0$ and $A^0$, we now have to consider the time-reversal transformation of the corresponding fields. 

\section{$CP$ symmetry}
In the SM, the scalar potential automatically preserves $CP$-invariance. Such is not the case in the 2HDM. For the sake of simplicity, let us first impose the invariance of the potential (\ref{potcust}) under the standard $CP$ transformation
\begin{equation}
\label{restrictedcp}
 (\mathcal{CP})\phi_i(t,\vec{r})(\mathcal{CP})^\dag={\phi_i^*}(t,-\vec{r})\quad (i=1,2).
\end{equation}
The operators $\hat{A}$, $\hat{B}$ and $\hat{C}$ are even under this transformation while $\hat{D}$ is odd. Consequently the custodial symmetry may be implemented in two different ways:
\begin{itemize}
 \item If $\gamma=0$, one has $\hat{C}'=\hat{C}$ such that the full potential (\ref{potcust}) is automatically $CP$-invariant. The $H_3$ state appears to be $CP$-odd and is thus identified with the $A^0$.  The $H_1$ and $H_2$ are $CP$-even and linear combinations of the ($h^0$,$H^0$) mass eigenstates. This corresponds to the ``usual'' custodial case where $m_{H^\pm}^2=m_{A^0}^2$.

 \item If $\gamma=\pi$, one has $\hat{C}'=\hat{D}$ such that all the terms \textit{linear} in $\hat{C}'$ must be set to zero in (\ref{potcust}) to respect invariance under $CP$. As a consequence $\Lambda_5=0$ and the two-by-two mass matrix (\ref{matmass}) is diagonal. The $H_3$ state is now $CP$-even and is thus called $H^0$. The $H_1$ and $H_2$ are  $CP$ and mass eigenstates to be identified with $h^0$ and $A^0$, respectively. This corresponds to a twisted custodial case  where $m_{H^\pm}^2=m_{H^0}^2$. 
\end{itemize}

However, (\ref{restrictedcp}) is not the most general $CP$ transformation. As emphasized in \cite{Branco:1999fs}, the $CP$ symmetry action on scalar fields in any model containing more than one Higgs doublet is not univoquely defined, in contradistinction to the SM. For the 2HDM, this freedom is simply parametrized in terms of one arbitrary phase:
\begin{eqnarray}
\label{refCP}
 (\mathcal{CP})\phi_1(t,\vec{r})(\mathcal{CP})^\dag&=&{\phi_1^*}(t,-\vec{r})\nonumber\\
 (\mathcal{CP})\phi_2(t,\vec{r})(\mathcal{CP})^\dag&=&e^{i\xi}{\phi_2^*}(t,-\vec{r}).
\end{eqnarray}
The operators $\hat{A}$ and $\hat{B}$ remain even under this symmetry, no matter the value of $\xi$, while the orthogonal combinations $\cos(\frac{\xi}{2})\hat{C}+\sin(\frac{\xi}{2})\hat{D}$ and $-\sin(\frac{\xi}{2})\hat{C}+\cos(\frac{\xi}{2})\hat{D}$ are respectively even and odd. By comparing these expressions with (\ref{Cprime}) one easily concludes that the usual custodial scenario corresponds to the choice $\xi=\gamma$ while the twisted one requires $\xi=\gamma-\pi$. This proves that it is always possible to disentangle the two scenarii independently of the $CP$ phase convention. Notice that in the intermediate cases (i.e., $\xi\neq \gamma,\gamma-\pi$), the four terms containing $\hat{C}'$ disappear from (\ref{potcust}). The potential is then invariant under a larger $SO(4)\times SO(4)$ symmetry which is spontaneously broken into $SO(3)\times SO(4)$. As a consequence the four components of $\phi_2$ are degenerate in mass, as seen from the equations (\ref{degmass}) and (\ref{matmass}).

\section{Comments and Conclusion}
For the standard choice $\xi=0$, the twisted custodial-invariant potential reads
\begin{equation}
\label{potcust2}
\tilde{V}= -m_1 \hat{A}-m_2 \hat{B}+\Lambda_1\hat{A}^2+\Lambda_2\hat{B}^2+\Lambda_3\hat{D}^2+\Lambda_4\hat{A}\hat{B}
\end{equation}
with $\hat{A}$, $\hat{B}$ and $\hat{D}$ defined in (\ref{operators}). In the limit $\Lambda_1=\Lambda_2=\Lambda_4/2$, the rather peculiar potential given in \cite{Chankowski:2000an} can be recovered after a suitable reparametrization. 

One genuine feature of the twisted custodial scenario is the presence of an accidental $\mathbb{Z}_2$ symmetry acting as
\begin{equation}
 \phi_1\rightarrow \phi_1\quad\mathrm{and}\quad \phi_2\rightarrow -\phi_2
\end{equation}
in the scalar potential (\ref{potcust2}). In the Higgs basis, the vev of $\phi_2$ vanishes. So this discrete symmetry is left unbroken and could advantageously supersede the $CP$-invariance required on the scalar potential to bring interesting phenomenology. For illustration it would nicely reconcile two apparent features of the electroweak interactions, namely natural flavour conservation and explicit $CP$-violation in the Yukawa sector \cite{Branco:1985pf}, if all fermionic fields are even under $\mathbb{Z}_2$. Were this the case, the lightest neutral component of the $\phi_2$ doublet (i.e. $H^0$ or $A^0$) would be a candidate for cold dark matter (see for example the inert doublet model in \cite{Barbieri:2006dq,LopezHonorez:2006gr}). 

The twisted custodial scenario may also provide interesting phenomenology at colliders \cite{deVisscher}. In particular, $A^0$ is no longer forced to be close in mass to the charged $H^\pm$ and is no more subject to the LEP bound due to its $CP$ assignment. So it may be relatively light and produced via the exotic $H^\pm\rightarrow W^\pm A^0$ process. Moreover, here $h^0$ is defined to be the $CP$-even component of $\phi_1$. Contrary to what is usually assumed in 2HDM studies, it can thus be heavier than all the other Higgs bosons and have atypical $h^0\rightarrow A^0A^0,H^0H^0,H^+H^-$ decays.

To summarize, we have implemented a twisted custodial symmetry such that the usual mass relation $m_{H^\pm}^2=m_{A^0}^2$ is turned into $m_{H^\pm}^2=m_{H^0}^2$, providing the natural frame for a light $A^0$ within the 2HDM. Equivalently, the substitution of the $CP$-even $H^0$ for the $CP$-odd $A^0$ can be understood in terms of a twisted $CP$ symmetry acting on the Higgs field. It would therefore be interesting to extend this analysis to the case of $n$HDM where the arbitrary $CP$ phase is generalized to an $(n-1)$-by-$(n-1)$ unitary matrix.

\section{Acknowledgements}
We thank Fabio Maltoni for useful discussions. This work was supported by the Institut Interuniversitaire des Sciences Nucl\'eaires and by the Belgian Federal Office for Scientific, Technical and Cultural Affairs through the Interuniversity Attraction Pole P6/11. 

\bibliography{custodial}

\end{document}